\documentclass[aps,prl,twocolumn,showpacs,amsmath,amssymb,floatfix]{revtex4}
\usepackage{epsfig}
\usepackage{bm}
\def\l{\langle\!\langle}
\def\r{\rangle\!\rangle}

\begin{document}
\title{Decay of a metastable state activated by non-Gaussian noise:\\
A critical review of the generalized Kramers problem}
\author{Eugene V. Sukhorukov$^1$ and Andrew N. Jordan$^2$}
\affiliation{$^1$\
D\'{e}partment de Physique Th\'{e}orique, Universit\'{e} de Gen\`{e}ve,
CH-1211 Gen\`{e}ve 4, Switzerland\\
$^2$\ Department of Physics and Astronomy, University of Rochester, Rochester, New York 14627, USA}

\date{\today}

\begin{abstract}
We review the problem of the noise activated escape from a metastable state in the presence
of non-Gaussian noise, and present connections between various theoretical approaches.
We also respond to criticism of our work by Tom\'a\v{s} Novotn\'y [arXiv:0807.0387] concerning the weak damping limit.
The discrepancy between our results is linked to uncontrolled approximations made by Novotn\'y. 
\end{abstract}


\maketitle

{\em Introduction.}---
The problem of noise activated escape from a metastable state has been formulated
and solved by Kramers more than 50 years ago \cite{kramers}.
This important problem has many applications across different fields.
It has been addressed many times in the literature, and reviewed in many textbooks,
e.g.\ in \cite{master}. By now, this problem is well understood on the level 
of Gaussian noise. Recently, new interest in Kramers' problem has been stimulated
by experimental progress in the field of electronic on-chip detectors, which 
allows the measurement of small non-Gaussian effects in electronic noise. Kramers' problem
was reconsidered beyond Gaussian noise and formally solved in Refs.\ 
\cite{tobiska,ourprl,ourPRB} for the case of strong damping and in Ref.\ \cite{us} for 
the case of weak damping, where the number of variables is reduced. This problem 
has been applied in Refs.\ \cite{us,ankerhold,grabert} to the stochastic dynamics of 
the Josephson junction (JJ) threshold detector, where the escape rate from the 
supercurrent state of the JJ provides the information about the current noise. 

It has been shown that the third cumulant of current noise, which describes a
weak non-Gaussianity of a Markovian process, contributes to the asymmetry of the escape 
rate with respect to the current bias. Very recently, this phenomenon has become a point 
of controversy. In the paper ``Josephson junctions as threshold detectors of the full 
counting statistics: open issues'' \cite{novotny} Tom\'a\v{s} Novotn\'y  claims that his new theory 
of noise-activated escape makes a prediction concerning non-Gaussian effects which contradicts 
previous theoretical results \cite{us,grabert}.  The disagreement only concerns a 10\% difference 
in a detector parameter's coefficient, and therefore would not otherwise be of great concern.
However, Novotn\'y goes on to comment that ``...it is relevant from a purely conceptual point of view 
which one is actually correct since it should help with the identification of possible misconceptions 
hidden in the failed approach(es).''  The purpose of the present short paper is to 
clarify and resolve this controversy and to explain the connection between different theoretical 
schemes.  

Here we show that the Novotn\'y theory represents a particular case of the stochastic 
path integral (SPI) formulation introduced in the author's earlier paper \cite{us} and relies on an 
alternative method of solving the resulting Hamilton-Jacobi equation for the classical action. 
However, the presented solution is not correct, which explains the discrepancy with previous 
theoretical results.  The mistake is not simple, but concerns the subtle nature of the weak-damping limit
and an uncontrolled approximation made by Novotn\'y.  Rather than simply point to the mistake, 
we take this opportunity to review the physics, and show how the approaches used in several papers are related.

\begin{figure}[tb]
\epsfxsize=6.5cm
\epsfbox{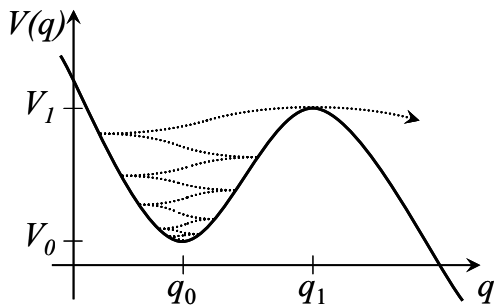}
\caption{A metastable system with the collective coordinate $q$ can be viewed as a ``particle'' 
which moves in a potential, shown by the full line, with a local minimum at the point $q=q_0$ 
and a maximum at the point $q=q_1$. The dotted line schematically shows the most likely trajectory 
which leads to the escape of the system from the metastable state. In the weak damping limit
the motion is quasi-periodic, and the number of winding increases with the quality factor $Q$
of the system.}
\vspace{0mm}
\label{trajectory}
\end{figure}

We start with the brief formulation of the problem. Consider a particle which moves in a 
metastable potential $V(q)$ around the local minimum at point $q=q_0$ (see Fig.\ \ref{trajectory}). 
Such a particle may represent a system in a state described by the collective variable $q$.
The conservative classical motion is generated by the Hamiltonian $H(p,q)=p^2/2m+V(q)$. In addition,
if the particle interacts with the environment, this leads to damping, so that the 
particle relaxes to the point at the minimum of the potential, $q=q_0$. At the same time,
the environmental noise activates the motion, so the particle may escape from the 
metastable state via the point $q=q_1$. We further assume that this complex behavior may be described 
by the set of Hamilton-Langevin (HL) equations,
\begin{equation}
\dot q=\partial H(p,q)/\partial p+I_q,\quad
\dot p=-\partial H(p,q)/\partial q+I_p,
\label{HLE}
\end{equation}
where the currents $I_q$ and $I_p$ are the sources of noise, which also describe damping. 
The problem is to find the rate of escape. 

{\em Formal solution of Kramers' problem}.---
If the currents are fast variables, so that they fluctuate on the time scale $\tau_0$,
which is much shorter than the the characteristic time scale $T$ of the deterministic
motion of the particle, then the HL equations can be solved by the method of the stochastic
path integral, introduced for non-Gaussian noise in Refs.\ \cite{SPI1,SPI2}. According to
this method, the evolution of the particle is described by the path integral 
\begin{equation}
P=\int\! {\cal D}\Lambda\!\int\!{\cal D}{\bf R}\exp(S),
\label{integral}
\end{equation} 
where the action $S$ is 
given in the explicitly canonically invariant form as
\begin{equation}
S=\int dt'[-\Lambda\cdot \dot {\bf R}+\Lambda\cdot \{{\bf R},H\}+
{\cal H}(\Lambda, {\bf R})].
\label{action1}
\end{equation}
Here ${\bf R}=(p,q)$ and $\Lambda=(\lambda_p,\lambda_q)$ are the sets of physical
and canonically conjugated auxiliary variables, respectively, $\{\ldots\}$
denotes the Poisson bracket with respect to $p$ and $q$, and the function ${\cal H}(\lambda_p,\lambda_q)$ 
is the generator of the cumulants of the Markovian noise sources:  
$\l I_p^nI_q^m\r=\partial^n_{\lambda_p}\partial^m_{\lambda_q}{\cal H}(0, {\bf R})$
and ${\cal H}(0, {\bf R})=0$.
By fixing ${\bf R}$ in the final state of (\ref{action1}) we obtain the 
probability distribution $P({\bf R})$, while fixing the final $\Lambda$ 
variables turns the SPI into the moment generating function $P(\Lambda)$.
That is why the auxiliary variables $\Lambda$ are also called ``counting''
variables in the field of full counting statistics \cite{Levitov}.

The separation of time scales $T\gg\tau_0$ leads to a small parameter $\tau_0/T$
which plays the role analogous to $\hbar$ in the Feynman path integral and thus
allows the saddle-point evaluation of the SPI. This naturally leads to Hamilton's equations of motion in the extended space \cite{SPI2}
\begin{equation}
\dot {\bf R}=\partial K/\partial\Lambda,\quad \dot\Lambda=-\partial K/\partial {\bf R},
\label{HE}
\end{equation}
generated by the new Hamiltonian 
\begin{equation}
K(\Lambda,{\bf R})\equiv \Lambda\cdot \{{\bf R},H\}+
{\cal H}(\Lambda, {\bf R})\,,
\label{K}
\end{equation}
which has to be solved for some initial condition ${\bf R}(t_0)={\bf R}_0$ and
$\Lambda(t_0)=\Lambda_0$ and substituted back to the action to give the 
evolution operator $P(t)=\exp[S(t)]$.

In the context of Kramers' problem there always exists a trivial solution $\Lambda=0$ and $\dot {\bf R}=\{{\bf R},H\}+\langle {\bf I}\rangle$, 
where ${\bf I}\equiv(I_p,I_q)$, for the ``average'' dynamics in physical space with a null Hamiltonian 
and action $S=K=0$, giving the proper normalization of the distribution $P$. This solution describes
the relaxation to the local minimum of the potential at $q=q_0$ (resting state), and away from the top 
of the barrier for $q\geq q_1$ (the running state). We are looking for a nontrivial solution 
with non-zero action which connects these two states and thus leads to the 
escape from the local minimum (it is shown in Fig.\ \ref{trajectory} by a dotted line). 
Therefore, the initial condition for this solution is 
$\Lambda(t_0)=0$, $p(t_0)=0$, and $q(t_0)=q_0$. 
Moreover, since the full Hamiltonian is an integral of motion,
and for the initial state of $K=0$,  
we have 
\begin{equation}
K[\Lambda(t),{\bf R}(t)]=0 
\label{integral2}
\end{equation} 
along the trajectory of interest. Therefore, the classical action simplifies and we find
\begin{equation}
S=-\int_{t_0}^{t_1} dt\Lambda(t)\cdot \dot{\bf R}(t)=-\int_{{\bf R}_0}^{{\bf R}_1}\Lambda({\bf R})\cdot d{\bf R}.
\label{action2}
\end{equation}
Then, up to a prefactor  \cite{footnote1}, the rate of the noise activated escape from the localized state 
is given by $\Gamma\propto\exp(S)$.

Although the original problem is stochastic, the advantage of the SPI method \cite{SPI2} is that it
reduces the problem to solving deterministic equations of motion in the extended phase space. Therefore,
Eqs.\ (\ref{HE}-\ref{action2}) provide a unique solution of Kramers' problem for an arbitrary
underlying conservative dynamics generated by $H(p,q)$ and for arbitrary Markovian noise generated
by ${\cal H}(\Lambda,{\bf R})$. However, these equations have a different equivalent representation,
which may have triggered the confusion in the paper \cite{novotny}. In the next part of the paper
we clarify the connection between these two representations. 

{\em Hamilton-Jacobi equation.}---
Following the textbook of Landau and Lifshitz \cite{LL}, we note that 
according to Eq.\ (\ref{action2}) the classical action
as a function of coordinates satisfies $\Lambda=-\partial S/\partial{\bf R}$.
We will use the symbol $S$ to represent both Hamilton's characteristic function (referred to as action), and
(when evaluated between the time limits on the escape trajectory) the exponential contribution to the activation rate,
as in (\ref{action2}).   Following this observation, equation (\ref{integral2}) can be written in the form
of a differential equation for the action,
\begin{equation}
K(-\partial S/\partial{\bf R},{\bf R})=0, 
\label{HJ}
\end{equation} 
which is nothing but the Hamilton-Jacobi (HJ) equation in the stationary case. A solution
of this equation in some region ${\bf R}\in\Omega$ requires fixing a boundary condition
at the edge $\partial\Omega$, which can be conveniently written as 
$\partial S/\partial{\bf R}|_{\partial\Omega}=-\Lambda_0$. There is nothing special 
about the direct solution of the equation (\ref{HJ}), if the correct physical boundary conditions
are satisfied. However, one should remember that such a solution is unique and should coincide
with the one that follows from the equations of motion. Indeed, one can equally consider 
the boundary condition for $\partial S/\partial{\bf R}$ as the initial condition for the canonical momentum
$\Lambda(t_0)=\Lambda_0$ and for the coordinate ${\bf R}(t_0)\in\partial\Omega$. 
Then the function $K$ generates the Hamiltonian dynamics with the canonical momentum $\Lambda({\bf R})$ 
which coincides with the vector field $-\partial S/\partial{\bf R}$
in the region $\Omega$. This procedure can be viewed as a method of characteristics
for solving the equation (\ref{HJ}).

The way Novotn\'y \cite{novotny} arrives at the HJ equation deserves a separate
consideration. Here we present the most general variant of it. Let us consider a 
specific noise process, one which is generated
by rare random transitions from the state ${\bf R}$ to the state ${\bf R}'$ with the rate 
$\Gamma({\bf R}',{\bf R})$. Since the transitions are rare (more rigorously, 
$\Gamma\tau_0\ll 1$), the probability distribution function $P({\bf R},t)$ satisfies 
the following master equation \cite{master}
\begin{eqnarray}
\dot P&=&\partial P/\partial {\bf R}\cdot \dot{\bf R}
+\partial P/\partial t=\{{\bf R},H\}\cdot\partial P/\partial {\bf R}\nonumber \\
&+&\int d{\bf R}'[\Gamma({\bf R},{\bf R}')P({\bf R}')-\Gamma({\bf R}',{\bf R})P({\bf R})],
\label{ME}
\end{eqnarray}
where the first term takes into account the deterministic part of the dynamics. 
We follow then the standard Kramers-Moyal forward expansion \cite{master} 
to derive the equivalent Fokker-Planck equation. The result is
\begin{equation}
\dot P=\{{\bf R},H\}\cdot\partial P/\partial {\bf R} +
{\cal H}(-\partial_{\bf R},{\bf R})P({\bf R}),
\label{FP}
\end{equation}
where the function ${\cal H}$ is defined as follows 
\begin{equation}
{\cal H}(\Lambda,{\bf R})=\int d{\bf R}_1\Gamma({\bf R}_1,{\bf R})\left[e^{({\bf R}_1-{\bf R})\Lambda}-1\right].
\label{H}
\end{equation}

In general, the order in which the derivative $\partial_{\bf R}$ is taken in (\ref{FP})
is important. However, accounting further for the separation of time scales 
parameter $T/\tau_0\gg 1$, and making the semiclassical approximation 
$P=\exp(S)$, we apply the derivatives
in (\ref{FP}) directly to the action. In this way in the stationary case $\dot P=0$
we again obtain the HJ equation (\ref{HJ})
with $K$ given by (\ref{K}) and the function ${\cal H}$ given by Eq.\ (\ref{H}). The paper  
\cite{novotny} then recognizes this function as the generator of the Poissonian process
and generalizes the result for an arbitrary noise. This step, although it gives the correct 
HJ equation, is not rigorous. In particular, this step cannot be directly applied   
to the Fokker-Planck equation (\ref{FP}), which simply does not exist for the 
case of general noise (or, if formally written, does not correspond to any physical 
reality). This is because a particular form of the FP equation fixes the quantization 
procedure via the operator ordering in (\ref{FP}), while the function ${\cal H}$ does
not provide enough information for this. 

Alternatively, on the SPI level the FP equation may be obtained 
from (\ref{integral}) in a standard way by expanding the ``one-step'' propagator 
$P({\bf R},{\bf R}';\Delta t)$ with respect to $\Delta t$. This step formally requires  
$\Delta t{\cal H}$ to be small. On the other hand, the SPI is constructed with the 
separation of time scales requirement, $T\gg\Delta t\gg\tau_0$, which inevitably leads
to the restriction ${\cal H}\tau_0\ll 1$. In turn, this implies that the noise is generated
by a weak Poissonian process, $\Gamma\tau_0\ll 1$, with the generator ({\ref{H}).  
However the SPI method itself does not rely on writing the FP equation as an intermediate step,
and therefore it is free from this difficulty and provides a rigorous and general way to describe 
the stochastic system on the semiclassical level. This fact is thoroughly explained in Ref.\ \cite{SPI2}.

{\em Energy diffusion}.---
To conclude the formal discussion of the problem, we consider the weak damping limit.
In this limit the system experiences a quasi-periodic motion (see Fig.\ \ref{trajectory}). 
The energy of the system weakly changes over the period making small steps back and forth. 
This process then can be viewed as diffusion of the energy, so that the action depends
only on the energy variable. For the rigorous derivation of the action via the canonical transformation of 
coordinates we refer the reader to our paper \cite{us}, and present here a simplified physical 
argument. 

The energy balance equation
\begin{equation} 
\dot E=\dot qI_p-\dot p I_q 
\label{balance}
\end{equation}
follows directly from the HL equations (\ref{HLE}) and takes the form of a Langevin equation.  
If $T_0\gg\tau_0$, then variables $\dot q$ and $\dot p$ change slowly and can be considered as 
``effective charges''. Therefore, the noise source on the right hand side of the equation 
(\ref{balance}) is described by the cumulant generating function ${\cal H}(\lambda_E\dot q,-\lambda_E\dot p)$.
Diffusion of the energy is then described by the SPI with the action 
\begin{equation}
S=\int dt'[-\lambda_E\dot E+{\cal H}(\lambda_E\dot q,-\lambda_E\dot p)].
\label{diffusion}
\end{equation} 
To leading order in weak damping we can replace the generator ${\cal H}$ in Eq.\ (\ref{diffusion})
with its average over the period $T$ of oscillations
\begin{equation}
\langle{\cal H}\rangle_E\equiv T^{-1}\oint dt {\cal H}(\lambda_E\dot q,-\lambda_E\dot p),
\label{average}
\end{equation} 
evaluated for fixed $\lambda_E$ and $E$.  Corrections in damping will be found by 
taking into account slow energy dissipation $\dot E=\partial_{\lambda_E}{\cal H}$ and  
$\dot \lambda_E=-\partial_E{\cal H}$, while averaging over the period $T$.
This gives the action
\begin{equation}
S=\int dt'(-\lambda_E\dot E+\langle{\cal H}\rangle_E).
\label{final-action}
\end{equation}

The simplification here arises from the fact that in contrast to the original action (\ref{action1}),
the Hamiltonian $K=\langle{\cal H}\rangle_E$ in the action (\ref{final-action}) depends on one variable 
only: the energy $E$. Therefore the condition $K=0$ becomes an algebraic equation for the instanton line
$\lambda_E=\lambda_{\rm in}(E)$, which connects the bottom and the top of the potential barrier, see Fig.\ 
\ref{trajectory}. Thus we arrive at the following result \cite{us}
\begin{equation}
S=-\int \lambda_{\rm in} dE,\quad \langle
{\cal H}(\lambda_{\rm in}\dot q,-\lambda_{\rm in}\dot p)\rangle_E=0,
\label{instanton}
\end{equation}
which formally solves the Kramers' problem for an arbitrary Markovian noise in the weak damping limit. 
This result will be compared with the alternative way of obtaining the weak damping limit.

{\em Josephson junction threshold detector}.---
The idea to use a Josephson junction in the metastable supecurrent state as a detector of noise 
belongs to Pekola \cite{pekola1} and to Tobiska and Nazarov \cite{tobiska}. Later on, the problem 
was reconsidered on a different level in the papers \cite{novotny,us,grabert,ankerhold}. Leaving 
the experimental realization \cite{pekola2,pekola3,saclay} of the idea aside, we focus solely on 
the theoretical part of the problem. For the physics of the JJ detector we refer 
the reader to our paper \cite{us}. Here we just mention that after rescaling the physical variables
\cite{novotny} the dynamics of the detector is described by the HL equations (\ref{HLE}) for a 
``particle'' with mass $m=1$ and with coordinate q (being a superconducting phase), 
which moves in the periodic potential biased by the supercurrent $J$: 
\begin{equation}
H(p,q)=p^2/2+V(q), \quad V(q)=-\cos(q)-Jq. 
\label{JJ-H}
\end{equation}

The dissipative part of the current localizes the JJ in the supercurrent state in one 
of the local minimums of the potential, e.g.\ at $q=q_0$, while the current noise 
activates the JJ and leads to the escape to the running dissipative state. This effect
enters via the momentum source $I_p$, while the coordinate source $I_q=0$. 
Truncating the noise generating function at the third cumulant, we write 
\begin{equation}
{\cal H}=\frac{1}{Q}(-p\lambda_p+\lambda_p^2)+c_3\lambda_p^3,
\label{JJ-generator}
\end{equation} 
where $Q$ is the quality factor of the JJ oscillator in the supercurrent state, and 
the constant $c_3$ is proportional to the third current cumulant, 
$\l I_p^3\r\equiv\partial^3_{\lambda_p}{\cal H}=6c_3$, which characterizes
a weak non-Gaussianity of noise. In all the papers on the subject $c_3$ 
is considered to be a small expansion parameter. 

Finally, the HJ equation (\ref{HJ}) acquires the following form:
\begin{equation}
-p\frac{\partial S}{\partial q}+V'(q)\frac{\partial S}{\partial p}
+\frac{p}{Q}\frac{\partial S}{\partial p}+\frac{1}{Q}\left(\frac{\partial S}{\partial p}\right)^2
=c_3\left(\frac{\partial S}{\partial p}\right)^3,
\label{JJ-HJ}
\end{equation}
where we deliberately neglect important physical effects in order to arrive at 
the HJ equation in the form presented in Ref.\ \cite{novotny}. Indeed,
in general the source $I_p$ depends on the system's state, which leads to 
the ``cascade corrections'' \cite{SPI2} to the action. In the context 
of the JJ physics, this is an effect of the measurement circuit backaction. 
Reference \cite{us} fully accounts for these circuit effects and comes to 
the conclusion that the corrections are small and can be neglected if the 
load circuit impedance is small. This, in fact, is the case in all existing 
experiments. 

{\em Weak non-Gaussianity}.---
We are now in the position to critically compare the results of Refs.\ 
\cite{novotny,us,grabert,ankerhold}. All of the papers implement a perturbation
expansion in the small parameter $c_3$ to leading order. We start with our original 
work \cite{us}, where the third cumulant contribution was found in the strong damping 
$Q\to 0$ and the weak damping $Q\to\infty$ limits.  We focus here on the 
weak damping limit and reduce the dynamics in the full space to the diffusion
in the energy space, as described above. Using Eqs.\ (\ref{instanton}) and 
(\ref{JJ-generator}), we solve the equation for the instanton line perturbatively
in $c_3$, $\lambda_{\rm in}=1 - c_3 Q\langle p^3\rangle_E/\langle p^2\rangle_E$, 
and arrive at the following result 
\begin{equation}
S=-\Delta V+c_3S_3,\quad S_3=Q\int_{V_0}^{V_1} dE\,\frac{\langle p^3\rangle_E}{\langle p^2\rangle_E},
\label{result1}
\end{equation}
where $\Delta V=V_1-V_0$ is the potential barrier, i.e.\ the first term is 
just Kramers' result for the Gaussian noise. Note that the second term 
contains a large prefactor $Q\to\infty$ which, on the other hand, is multiplied
by the small integral, because $\langle p^3\rangle_E=0$ to leading order in damping. 
This can be explained by the fact that for large $Q$ the escape trajectory (shown in
Fig.\ \ref{trajectory}) makes small energy steps and large number of windings. 
In the appendix of Ref.\ \cite{us} we take weak damping into account,
$\dot E=p^2/Q$ (as explained above), to find the next order correction to $\langle p^3\rangle_E$, and arrive at
\begin{equation}
S_3=2\int_{V_0}^{V_1} dE\frac{\langle (q'-q)p^2\rangle_E}{\langle p^2\rangle_E},
\label{result2}
\end{equation}
where $q'$ is one of the two turning points of the conservative motion, 
which is closest to the point $q_1$.
Thus we find that the quality factor $Q$ cancels from the final result. 

Grabert's paper \cite{grabert}, despite being long, can be summarized in just one paragraph.
It utilizes the SPI method of Ref.\ \cite{us} and directly solves Eqs.\ 
(\ref{HE}-\ref{action2}) for arbitrary damping. On the Gaussian level
the escape path that satisfies Hamilton's equations of motion and all 
the boundary conditions listed before Eq.~(\ref{integral2}) is described by the equations 
\begin{equation}
\dot q=p,\quad \lambda_q=V',\quad \dot p=-V'+p/Q,\quad \lambda_p=p,
\label{escape}
\end{equation}
which is time reversed with respect to the corresponding relaxation path. Substituting these equations
into the action (\ref{action2}), we obtain the Gaussian part of the action,
\begin{equation}
S_{\rm G}(p,q)=-H(p,q)=-p^2/2-V(q),
\label{action-G}
\end{equation} 
which when evaluated between $t_0$ and $t_1$ along the escape trajectory gives Kramers' result $S_G = -\Delta V$.
Next, considering the term ${\cal H}'=c_3\lambda_p^3$ in (\ref{JJ-generator}) as a perturbation, we note that to leading order in $c_3$, the variation 
of the trajectory does not contribute to the action (\ref{action1}), so $S=S_{\rm G}+\int dt {\cal H}'$. 
Therefore, using again $\lambda_p=p$ we obtain that the total action takes the 
form (\ref{result1}) with
\begin{equation}
S_3=\int_{t_0}^{t_1} dt [p(t)]^3,
\label{result3}
\end{equation} 
where the integral is taken over the escape path. This is
the main result of the Ref.\ \cite{grabert}, which generalizes the calculations of
Ref.\ \cite{us} to arbitrary damping $Q$. In the weak damping limit, one easily 
obtains Eq.\ (\ref{result1}). Indeed, we can write $\int dt p^3=\int dt \langle p^3\rangle_E
=\int dE \langle p^3\rangle_E/\dot E=Q\int dE \langle p^3\rangle_E/\langle p^2\rangle_E$, where
we have used $\dot E=p^2/Q$. 

Finally, the Refs.\ \cite{novotny,ankerhold} choose to solve the HJ equation (\ref{JJ-HJ})
using different approximations. Here we reproduce a few steps of Ref.\ \cite{novotny}.
To leading order in the small parameter $c_3$ one can write $S=S_{\rm G}+c_3S_3$. Substituting
this expansion to the equation (\ref{JJ-HJ}) and using the Kramers action (\ref{action-G}), 
one arrives at the following equation:
\begin{equation}
p\frac{\partial S_3}{\partial q}-V'(q)\frac{\partial S_3}{\partial p}
+\frac{p}{Q}\frac{\partial S_3}{\partial p}=p^3.
\label{novotny-eq}
\end{equation} 
This equation can be solved exactly using the method of characteristics by choosing
$S_3[p(t),q(t)]$ with $p(t)$ and $q(t)$ being solutions of the equations (\ref{escape}) and describing the escape path in Fig.\ \ref{trajectory}.
In this case, equation (\ref{novotny-eq}) may be rewritten as $dS_3/dt=p^3$ by using equations (\ref{escape}), which recovers the previous result
(\ref{result3}). Therefore, the direct solution of the HJ equation
gives exactly the same result as solving Hamilton's equations of motion, as expected. The mistake 
of Ref.\ \cite{novotny} is that it treats the limit $Q\to\infty$ incorrectly
by dropping the $1/Q$-term in the above equation, giving Novoton\'y's equation
$p \partial_q S_3-V'(q)\partial_pS_3=p^3$.   This step would be equivalent to dropping the
$1/Q$ term in the equations of motion (\ref{escape}) in our approach.  Indeed, the dropping
of this term corresponds to an exactly constant energy, ${\dot E} =0$, rather than the physically
correct slowly growing energy, ${\dot E} = p^2/Q$.   After this uncontrolled approximation, 
the Novotn\'y equation may be integrated
over the energy conserving periodic trajectories, leading to 
$S_3=\int dt \dot qp^2$. After integrating by parts and using $\dot p=-V'$
one arrives at $S_3=(q_1-q_0)p^2+2\int dq V'(q)q$, which is the expression
(9) of the Ref.\ \cite{novotny}. However, this solution corresponds to exactly energy-conserving trajectories, and therefore
cannot be correct.  This explains why Ref.\ \cite{novotny}
finds the parametrically correct contribution of the third cumulant, while the 
dimensionless function of the current bias $D_1(J)$ is wrong in the detailed dependence on $J$. 

To summarize, we have reviewed the generalized Kramers problem of the 
decay of a metastable state under the influence of the non-Gaussian noise. 
We compared different calculations and came to the conclusion 
that they utilize basically the same theoretical framework introduced
earlier in the papers \cite{SPI1,SPI2}. However, Refs.\ \cite{novotny,ankerhold} make uncontrolled approximations in
solving Hamilton-Jacobi equations and thus arrive at parametrically correct, but quantitatively wrong results.

\bibliographystyle{apsrev}

\begin{thebibliography}{99}
\bibitem{footnote1} The prefactor in $\Gamma$ remains unknown, 
because it depends on non-Markovian behavior at short times 
of order $\tau_0$.
\end{thebibliography}

\end{document}